# EpidermaQuant: Unsupervised detection and quantification of epidermal differentiation markers on H-DAB-stained images of reconstructed human epidermis

Dawid Zamojski[1,2], Agnieszka Gogler[3], Dorota Scieglinska[3], Michal Marczyk[1,4,*]

[1]Department of Data Science and Engineering, Silesian University of Technology, Gliwice, Poland

[2]Genetic Laboratory, Gyncentrum Company, Sosnowiec, Poland

[3]Centre for Translational Research and Molecular Biology of Cancer,
Maria Sklodowska-Curie National Research Institute of Oncology, Gliwice, Poland

[4]Yale Cancer Center, Yale School of Medicine, New Haven, CT, USA

**\*Corresponding author:**

Michal Marczyk

Department of Data Science and Engineering, Silesian University of Technology

Akademicka 16, 44-100 Gliwice, Poland

E-mail: michal.marczyk@polsl.pl

**Abstract**

**Background:** The integrity of the reconstructed human epidermis generated *in vitro* could be assessed using histological analyses combined with immunohistochemical staining of keratinocyte differentiation markers. Computer-based analysis of scanned tissue saves the expert time and may improve the accuracy of quantification by eliminating interrater reliability issues. However, technical differences during the preparation and capture of stained images and the presence of multiple artifacts may influence the outcome of computational methods. Also, due to the specific nature of the analyzed material, no annotated datasets or dedicated methods are publicly available.

**Results:** Using a dataset with 598 unannotated images showing cross-sections of *in vitro* reconstructed human epidermis stained with DAB-based immunohistochemistry reaction to visualize 4 different keratinocyte differentiation marker proteins (filaggrin, keratin 10, Ki67, HSPA2) and counterstained with hematoxylin, we developed an unsupervised method for the detection and quantification of immunohistochemical staining. The proposed pipeline includes the following steps: (i) color normalization to reduce the variability of pixel intensity values in different samples; (ii) color deconvolution to acquire color channels of the stains used; (iii) morphological operations to find the background area of the image; (iv) automatic image rotation; and (v) finding markers of human epidermal differentiation with clustering. Additionally, we created a method to exclude images without DAB-stained areas. The most effective combination of methods includes: (i) Reinhard's normalization; (ii) Ruifrok and Johnston color deconvolution method; (iii) proposed image rotation method based on boundary distribution of image intensity; (iv) k-means clustering using DAB stain intensity.

**Conclusions:** The results of the present work should enhance the performance of quantitative analysis of protein markers in reconstructed human epidermis samples and enable comparison of their spatial distribution between different experimental conditions.

The algorithm is publicly available here: https://github.com/DawZam/EpidermaQuant.

**Keywords**

## Background

Immunohistochemistry (IHC) allows to visualize specific proteins in fixed tissue sections. It involves the application of antibody specific to a target antigen, which ultimately allows microscopic observation of the antigen-antibody complex in situ [1]. The so-called secondary antibody that can be conjugated to an enzyme, such as horseradish peroxidase (HRP) is used for visualization of antigen-antibody binding in DAB (3,3'-Diaminobenzidine)-mediated IHC staining. HRP oxidizes DAB and converts it into an insoluble brown precipitate [1]. The presence and localization of brown staining in tissue can be evaluated and imaged using light microscope [2]. The use of quantitative digital analysis of images obtained by IHC staining is crucial for researchers in the field of pathology [3]. It can facilitate the development of novel diagnostic methods and can provide additional information relevant to diagnosis and subsequent treatment. It is also important for basic research, e.g. it can provide quantitative assessment of protein markers in tissue-like (organotypic) cultures generated using 3D *in vitro* cell culture techniques. For example, the *in vitro* reconstructed human epidermis (RHE) is considered a representative and sufficient model of the human epidermis. RHEs can be obtained by culturing keratinocytes at the air-liquid interface on a collagen-fibroblast matrix. Such a 3D *in vitro* cell culture is a valuable tool for studying the impact of selected gene products on the integrity and function of the human epidermis [3].

The preparation of IHC slides, the method of image acquisition, technical differences during image capturing, and the presence of various artifacts are the main challenges in computational detection and segmentation of DAB-stained areas. The solution to this problem is to use an unsupervised algorithm for processing IHC images. Compared to the frequently used hematoxylin and eosin-stained samples, the colors of DAB-stained samples are less consistent, which is considered a major problem in fully automated image processing systems. In addition, a multitude of obstacles and limitations make it difficult for both the specialist and the algorithm to produce consistent results [4]. The variability in shape, size, and color of interesting objects, the lack of clear criteria for cell type identification, and the presence of overlapping structures should be mentioned as the main difficulties. An effective image pre-processing concept should overcome all these limitations to ensure high quality and accuracy in the studies performed, especially clinical ones. With the help of computer-assisted evaluation of IHC-stained digital tissue (such as bioptates) images, a quick and more accurate prognosis or a new understanding of the mechanism of the disease is possible.

A variety of techniques have been proposed to segment and quantify tissue stained with hematoxylin-eosin on digital images [5]. Studies related to DAB staining are still limited. Patel et al. used the Otsu method for negative control slides to define the threshold that distinguishes tissue from the background, and all pixels considered tissue were evaluated for normalized red minus blue (NRMB) color intensity. Then, a user-defined error tolerance was applied to the negative control slides to set the NRMB threshold, distinguishing DAB-stained from unstained tissue. This threshold was used to calculate the pixel fraction of DAB-stained tissue on each test slide [6]. Bencze et al. compared semi-quantitative scoring and digital analysis of DAB-stained images using a Convolutional Neural Network. Semi-quantitative scoring was carried out using a four-point scale based on the IHC intensity of the cells. Users might submit images, which were then analyzed to determine whether the algorithm had correctly recognized the objects or if manual annotation of points of interest was required [7]. Eszter Sziva et al. proposed a mechanism for quantitative histomorphometric-mathematical image analysis of DAB-stained tissue as support in andrology and reproductive medicine. The quantitative analysis was based on the use of features such as area, circumference, and cross-sectional diameter of stained testes [8]. In their work, Roszkowiak et al. described a comprehensive system capable of quantifying digitized DAB&H-stained breast cancer tissue samples at different intensities. The researchers distinguished segmentation based on regions of interest, as well as a step of dividing a cluster of nuclei, followed by boundary enhancement. The system used machine learning and local recursive processing to eliminate distorted and inaccurate regions of interest contours [4]. Deep learning methods are gradually overtaking classical image processing methods that deal with feature space partitioning. This is particularly evident in images where the features of objects are uncertain and variable, as is the case of IHC-stained tissue images [9], [10]. Some IHC image analysis methods have been developed and made available as freeware, for example, *QuPath* or *ImageJ* software [11], [12].

Page **5** of **20**

Here, we developed an unsupervised method for detecting protein markers of human epidermal keratinocyte differentiation on IHC images. The need to develop a new method was mainly due to: (i) technological limitations of current IHC image processing methods, which are not dedicated to epidermal images; (ii) problems with the structural complexity of the 3D tissue-like models resulting from their culture methodology; (iii) significant variability in the structure and shape of the areas of interest between the analyzed images, resulting from the specificity of the cell growth in the post-confluent cell density. We tested the effectiveness of various methods for pre-processing and segmentation of IHC-stained epidermal differentiation markers, proposed our solutions, and summarized the best methods into a freely available pipeline.

**Methods**

*Data*

The dataset consisted of 598 unannotated .jpg images (1936x1460 pixels) containing formalin-fixed and paraffin-embedded sections of *in vitro* reconstructed human epidermis, tissue-like structures generated by 3D culture *in vitro* of immortalized epidermal keratinocyte HaCaT line. Samples were processed by immunohistochemistry and subsequently stained using antigen-specific primary antibody and appropriate secondary antibody. After visualization of antigen-antibody binding by DAB-mediated reaction, cell nuclei were counterstained with hematoxylin (see examples in **Figure 1**). The details on tissue processing and IHC protocols are described in Gogler-Piglowska et al. [13]. Nine biological repeats in the form of positive control (specimens that contained the target molecule in its expected location and enabled the visualization of histomorphology) and negative control without primary antibodies (to confirm the specificity of the IHC staining) were prepared [14]. Four selected proteins (filaggrin (FLG), keratin 10 (K10), Ki67 protein (Ki67), and HSPA2 protein (HSPA2)) considered as markers of keratinocyte differentiation which are present in individual layers of the epidermis, indicative of epidermal development, were analyzed. The number of images per marker was as follows: FLG - 189 images (163 positive and 26 negative controls); K10 - 131 images (113 positive and 18 negative controls); Ki67 - 96 images (96 positive and 0 negative controls); HSPA2 - 182 images (124 positive and 58 negative controls). Methods were tested using images of FLG, K10, and Ki67 markers and the final pipeline was visually tested using HSPA2 marker data. Images were captured using a ZEISS AXIOCAM 503 color camera with a ZEN 2.6 photo archiving system at 40x magnification. Data preparation and scanning was done at the Centre for Translational Research and Molecular Biology of Cancer, Maria Sklodowska-Curie National Research Institute of Oncology, Gliwice Branch, Poland.

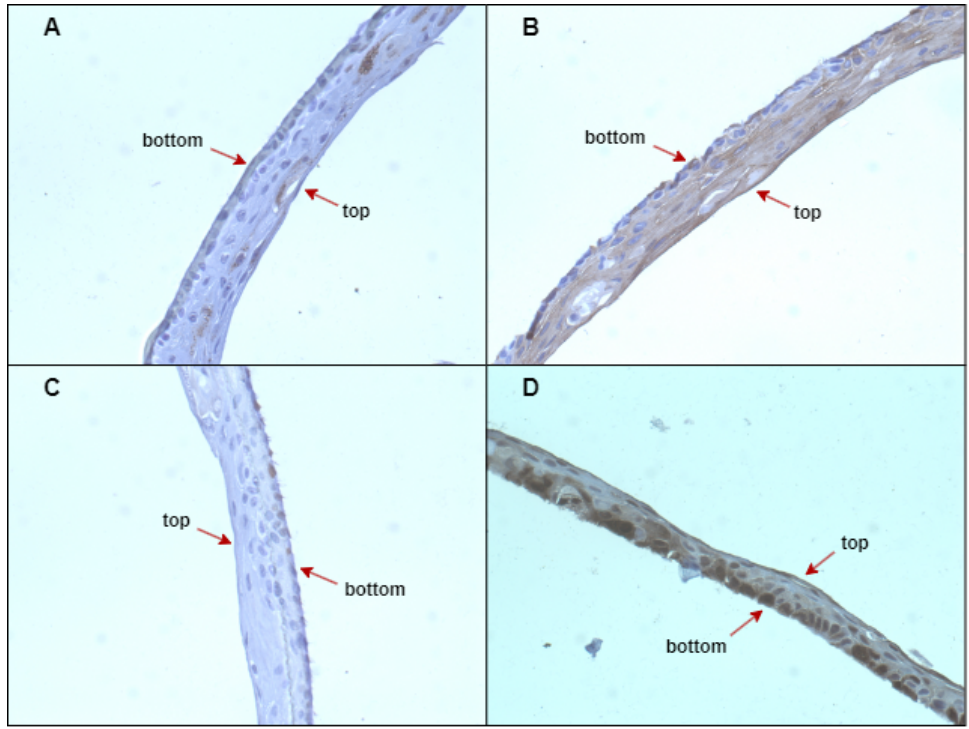

*Figure 1 An example of IHC images of reconstructed human epidermis stained for:* A - FLG marker; B - K10 marker; C - Ki67 marker; D - HSPA2 marker.

*Color normalization and deconvolution*

Reinhard's method, normalization by RGB histogram specification, and Macenko's method were tested as the implementation of the non-linear staining normalization algorithms [15]. Reinhard's method involved applying the *mat2gray()* transformation function to improve results. This operation converts a matrix to a grayscale image and stretches the histogram intensity values from 0 to 1. Then, to transform stained biological samples into images representing the concentrations of individual stains, color deconvolution was performed [16]. The method by Ruifrok and Johnston with different sets of parameters was used: (i) from the *ImageJ Color Deconvolution 2 plug-in*; and (ii) from the *Python scikit-image* implementation (see details in the **Supplementary Text**) [17].

*Background detection and mask creation*

The image on which Otsu's binarization method was performed was a greyscale hematoxylin channel. Morphological operations were implemented to remove noise and unnecessary regions in the specimen including the image background. Morphological operations were applied in the following sequence: opening, closing, and hole filling. Opening removed unconnected light objects that are smaller than the structuring element (SE), while closing removed unconnected dark objects that are smaller than the SE. The SE was a disk with a radius of 15. Filling holes operation completed the background, that is the empty spaces between pixels close to each other. Subsequently, artifacts were discarded by selecting the largest area of the tissue slide on a previously created black and white mask. In the end, the morphological operations were used once again to improve the quality of the results.

*Image rotation and crop*

Two methods of image rotation were tested: (i) the Hough transform; and (ii) the proposed image rotation method based on the boundary distributions of pixel intensities.

The principle of the Hough transform is based on the spatially transformed points so that the lines in the image are represented by points in Hough space. Then, by analyzing the relationships between the points in Hough space, the algorithm can identify the lines in the image. The Hough space contains parameters that describe the shape being searched for. For example, to detect a straight line in an image, the Hough space is two-dimensional and consists of the angle of the line and the distance of that line from the origin of the coordinate system [18].

The proposed image rotation method automatically rotates the tissue to a horizontal position based on maximizing the boundary pixel intensities. The boundary intensity distribution is calculated as the sum of the intensities of the pixels in each row. When the orientation of the tissue is horizontal, the intensity distribution reaches maximum values, which allows the algorithm to accurately arrange the tissue to a horizontal position. The proposed algorithm is summarized in the following steps:

- the tissue mask (negation of background mask) is used as the input,

- the mask is rotated from 1 to 180 degrees with a step of 0.1 degrees,
- for each rotation position, the sum of pixels in each row of the mask is calculated to create boundary intensity distribution,
- for each rotation position, the maximum value of the boundary intensity distribution is found,
- the final rotation position is the one that maximizes the maximum value of the boundary intensity.

*Elimination of images without DAB stain*

To prevent the segmentation of images without DAB staining, an average proportion (AP) measure was introduced. The determination of minimum staining level was carried out for a set of images representing the FLG marker as the most difficult to analyze. The result for an individual DAB channel image was expressed as a percentage of the number of pixels passing the condition to the total number of pixels in the image:

$$AP = \frac{no.\ of\ pixels\ <\ threshold}{total\ number\ of\ pixels} * 100\ [\%] \quad (1)$$

To determine the appropriate threshold and the value of the average proportion, which is a determinant of the elimination of images without DAB stain, ROC curve analysis was performed and the Youden index was calculated. The threshold for AP was searched in the range from 0.1% to 1% with a step of 0.01.

*DAB region segmentation*

Segmentation algorithms in the context of IHC imaging can be used to find areas stained with particular stains. From the existing segmentation methods, the k-means clustering algorithm was selected due to its efficiency and simplicity [19]. The process was divided into several steps: (i) reshape the DAB image into a new matrix retaining the original data; (ii) normalize DAB image channels independently using the mean and standard deviation; (iii) estimation of the number of clusters using the Davies-Bouldin method, based on the ratio of distances within a cluster and between clusters, in the range from 1 to 3; (iv) k-means clustering; (v) mask DAB image using obtained clusters; (vi) arrange clusters from the darkest using average DAB intensity per cluster.

*EpidermaQuant framework*

The final pipeline for analysis was based on the custom algorithm written in MATLAB® R2021b programming environment (see **Supplementary text**; **Supplementary Figure 1**). The algorithm is publicly available here: https://github.com/DawZam/EpidermaQuant.

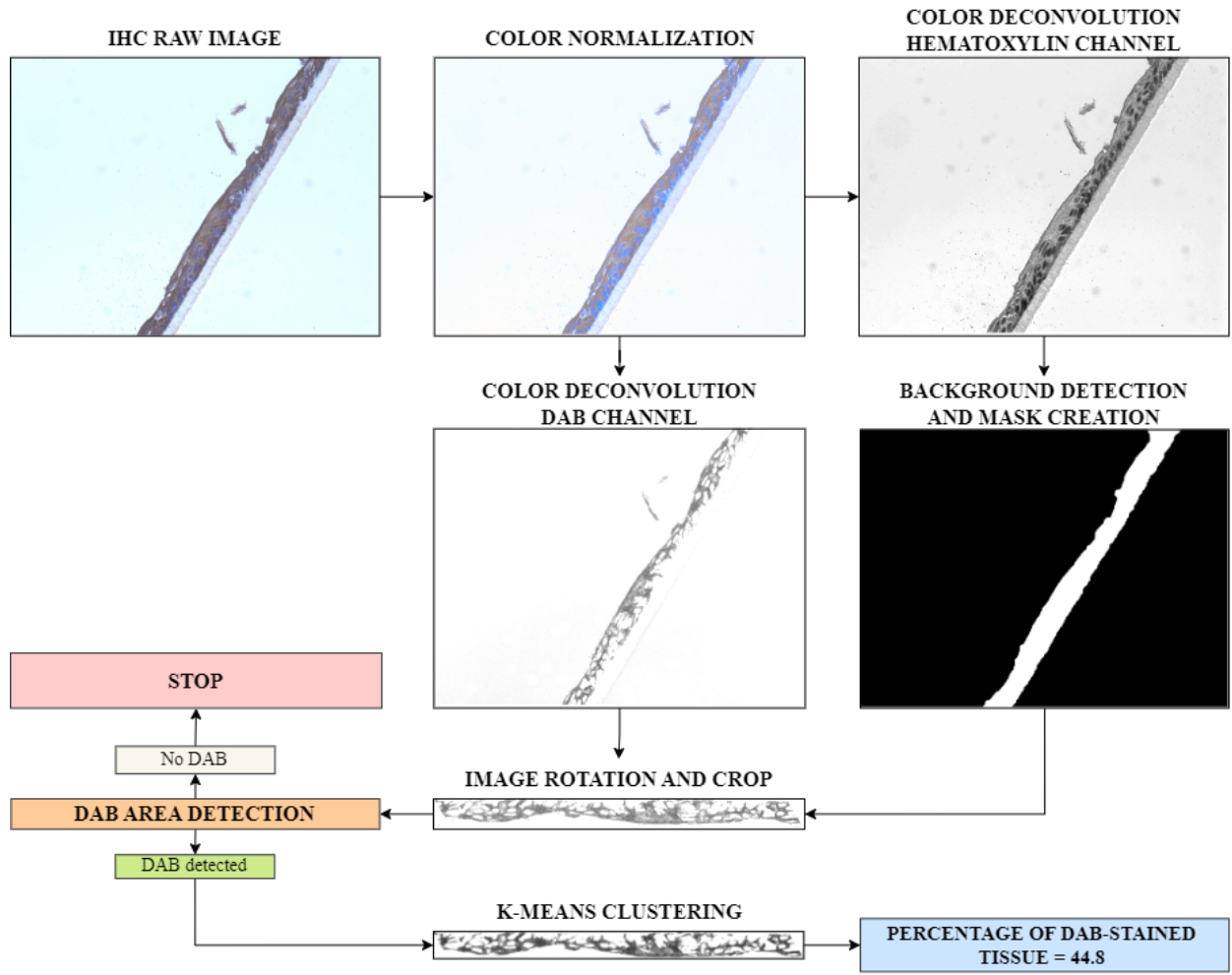

*Figure 2 EpidermaQuant algorithm flowchart. The main steps of the pipeline include image color normalization, image color deconvolution, background detection, image rotation and crop, initial DAB area detection step where the algorithm decides whether the image should be further analyzed, image segmentation, and calculation of the DAB percentage on tissue.*

**Results**

The analyzed pre-processing steps include image normalization, color deconvolution, background removal, and image rotation. Then, the image is segmented using the k-means clustering method. We tested selected methods and the best algorithms chosen by visual investigation of resulting images by experts were implemented in the EpidermaQuant framework (**Figure 2**).

*Color normalization and deconvolution*

To better distinguish the interesting objects in the image and harmonize images from different scans, we tested several color normalization and color deconvolution methods (**Supplementary Figure 2**). The best normalization method was Reinhard's normalization. In most images, the brown staining was best exposed and background noise was reduced in comparison to others (**Figure 3A**). Macenko's method gave visually poorest results. The effectiveness of the color deconvolution method was tested using the results of the hematoxylin and DAB channels (**Figure 3B**). The best set of parameters was from the *ImageJ Color Deconvolution 2 plug-in.* Only this method provided proper extraction of the DAB image including only areas with brown staining representing the analyzed marker. Also, the residual image showed low intensities, which is preferential. In other methods, background noise was visible in the DAB image (**Figure 3B**).

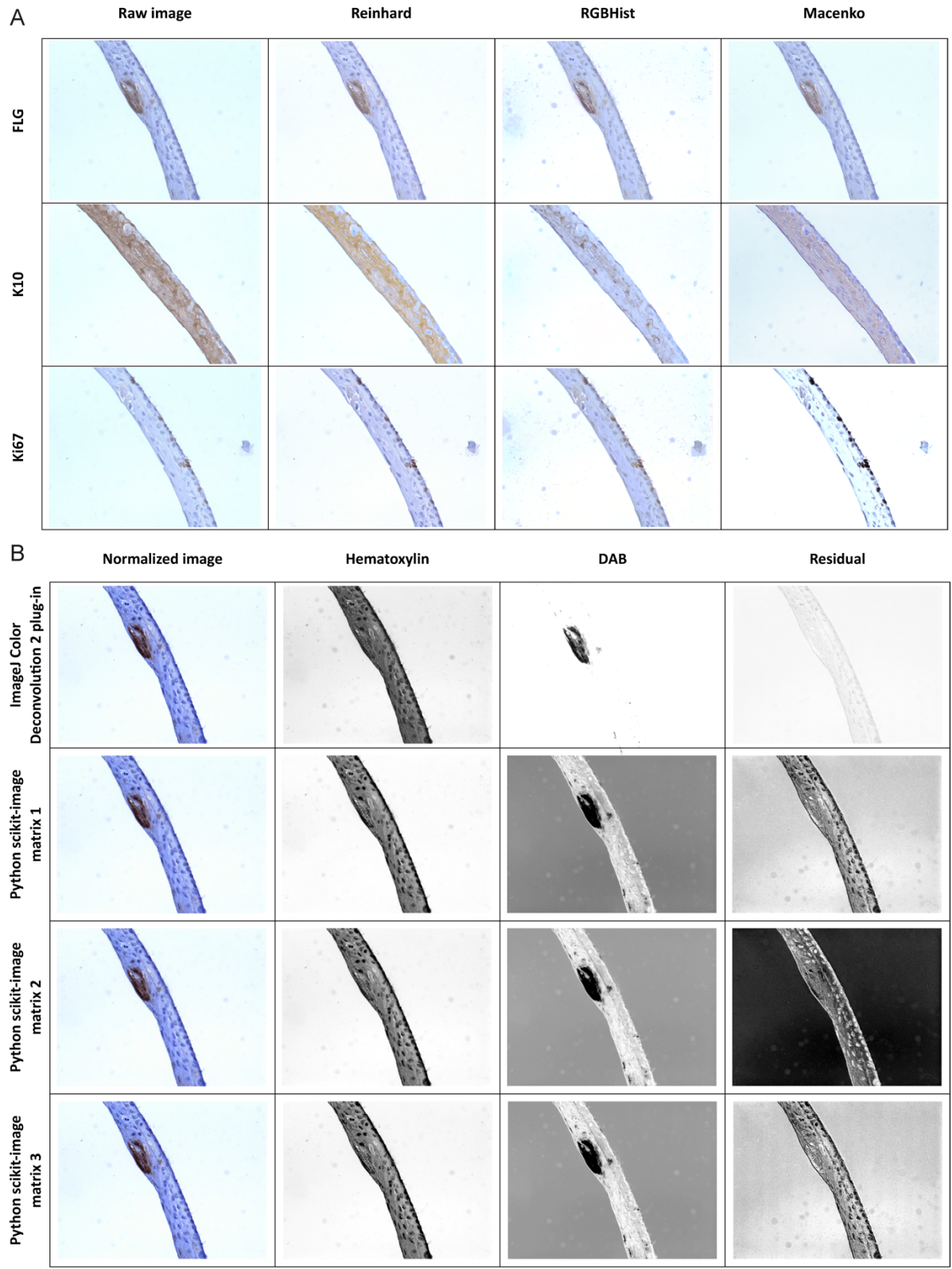

Figure 3 Comparison of normalization (A) and color deconvolution (B) methods with the original image depending on the selected marker using exemplary images.

*Background removal and image rotation*

We detect a background region using the image of the hematoxylin channel since most tissue structures are visible here. The first step of the algorithm, image binarization, gives a rough estimate of tissue regions (**Figure 4**). By introducing a series of subsequent morphological operations and binarization we find the background area of the image and create a mask without artifacts. Lastly, we discard small artifacts by selecting the largest area of the mask. The mask allows the definition of a clear boundary between a tissue slice and an unnecessary image background.

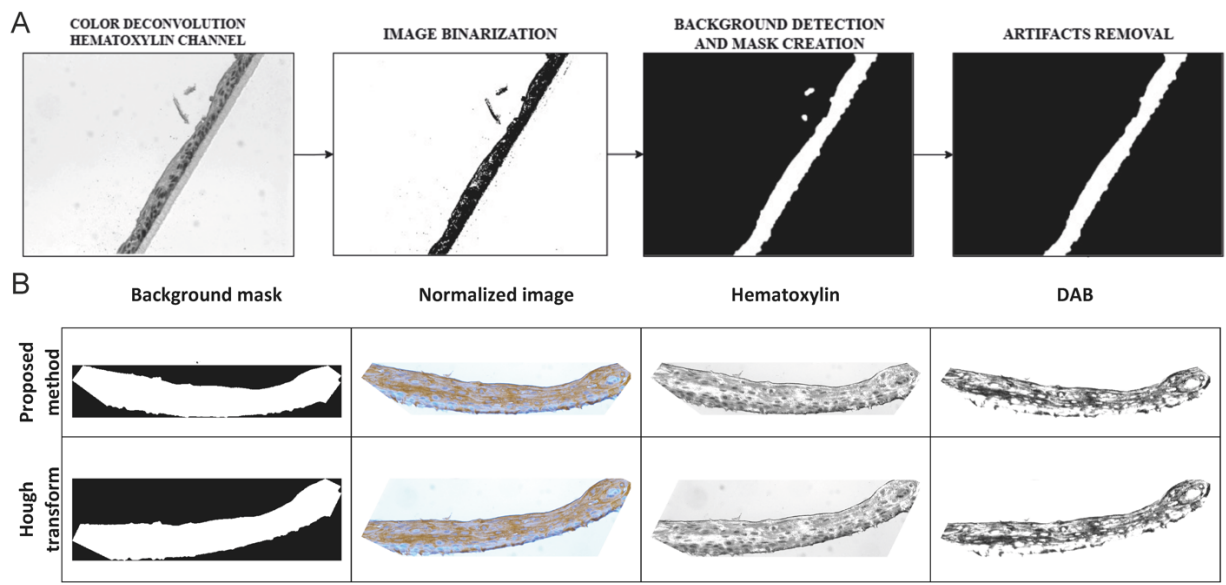

Figure 4 Subsequent steps of background removal (A) and comparison of two image rotation methods for an exemplary image of K10 marker (B).

The reconstructed epidermis forms a thin, single stripe on each stained slide. To reduce scanned image size, we rotate the image and crop as much background as possible (**Supplementary Figure 3**). At first, we tested functions implementing the Hough transform, however, its performance was not effective when tests were conducted on the entire set of images. Therefore, we developed the image rotation method based on the boundary distribution of image intensity, which proved to work much more efficiently (**Figure 4B**). Based on the results of the mask rotation, we crop the DAB channel image for further analysis.

*Elimination of images without DAB stain*

Some tissues may not contain the analyzed marker, so no DAB-stained regions should be observed on the image. Since we noticed brown-colored individual pixels in the negative samples, we developed a method to find and remove negative samples from further analysis. First, we calculated the AP for all images representing the FLG marker in both, positive (with marker) and negative samples. By maximizing the AUC and Youden indices, we found the optimal threshold value as 0.6% (**Supplementary Figure 4**).

*DAB area detection*

Using the DAB channel of the image, we found areas of interest in the form of DAB staining in all images representing each marker. In most cases, 2 clusters were obtained using the k-means algorithm (**Figure 5**). The cluster containing the DAB image area with the largest number of darkest pixels forms the final segmentation of the analyzed marker region. These outcomes were used to analyze the results quantitatively, to determine the percentage of DAB-stained tissue, and to mark the outlines of the DAB areas on the original image.

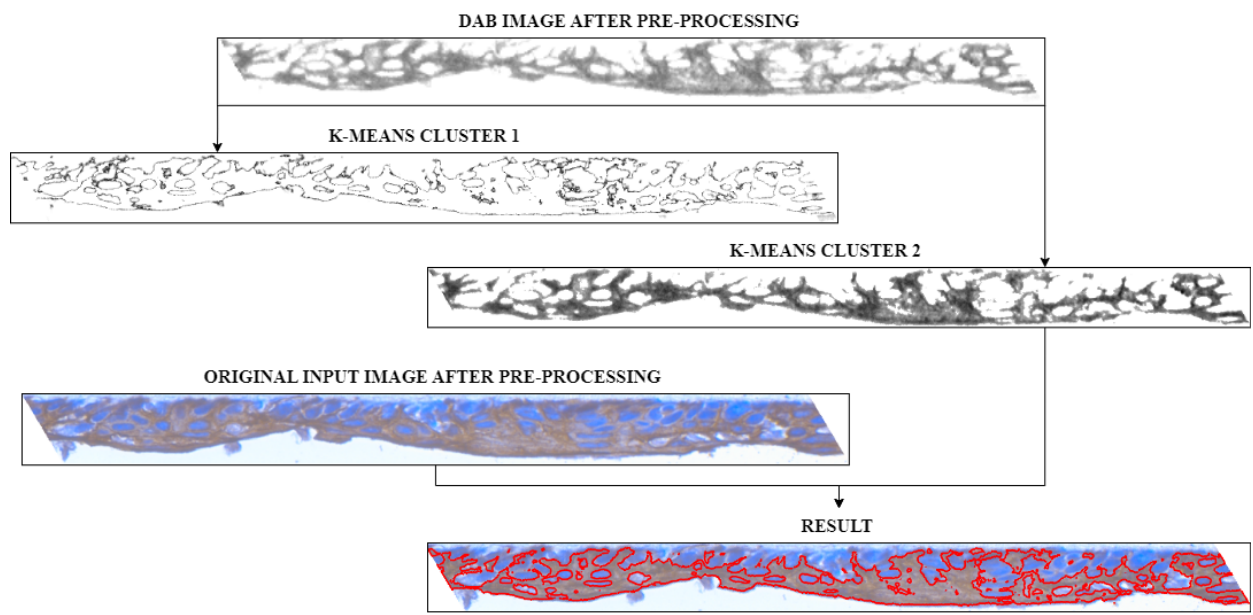

*Figure 5 Example of DAB image segmentation result.* Interesting regions of markers of human epidermal differentiation are segmented by the k-means method. The selected cluster outlines the DAB-stained areas on the original image.

*Comparison of DAB-staining levels between markers*

The percentage of DAB-stained tissue area for the representatives of each study group is compared in **Figure 6**. Samples representing the K10 marker showed the biggest area with DAB stain in comparison to other markers (median equal to 39,5% vs 4,1% in FLG, 2,5% in Ki67, and 25,3% in HSPA2). The spread of results obtained was also the largest. The smallest spread of results and, at the same time, the smallest area of the DAB stain were found for the Ki67 marker (see examples in **Supplementary Figures 5,6,7, and 8**).

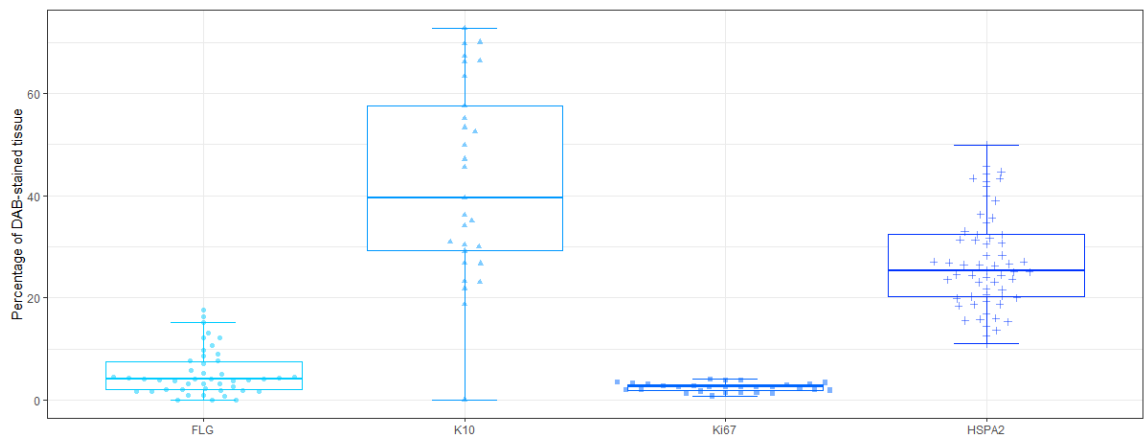

*Figure 6 The percentage of DAB-stained tissue. The graph shows the percentage of DAB occupancy on the slide in each sample by study marker group.*

## Discussion

We developed a new pipeline for automated detection, segmentation, and quantitative analysis of human epidermis differentiation markers in images containing slides of *in vitro* reconstructed human epidermis stained by DAB-mediated IHC and counterstained with hematoxylin. A performed analysis led to the choice of the best methods in subsequent processing steps.

Stain color inconstancy is a known problem, that could be caused by several factors: the thickness of the tissue section, antibodies concentration, stain timings, stain reactivity, characteristics of the scanner, and others. For the color normalization step, Reinhard's normalization was chosen because its results best reflected the colors of the staining used, especially for the brown areas of the DAB. The other two methods produced less satisfactory results. Both the RGB histogram specification and Macenko's normalization methods failed to sufficiently highlight areas of interest in the form of DAB staining. Our results were dissimilar from the study conducted by Khan et al. [15], where the authors selected Macenko's normalization method as the best. They mentioned that Reinhard's method was based on the false assumption of unimodal color distribution in each channel as the reason for rejecting it. This confirms that there are no gold standards and tests should be performed on each different tissue and staining type.

In the case of the image rotation method, it was decided to develop a new method that was more efficient compared to the Hough transform. In most cases it excluded situations in which the image was not rotated maximally to a horizontal position or when it was incorrectly rotated to a vertical position. This step was important for the proper presentation of the segmentation results. Without it, it would have been difficult to compare images in different positions with each other, especially in the case of images representing the FLG marker, where the sizes of clusters of DAB staining areas are very small.

The analyzed images were mostly divided into two or three clusters, correctly reflecting the areas of interest. The k-means algorithm was also used in a paper by Vidhya et al, which presented an effective technique for segmenting and classifying IHC tissue images to detect colorectal tumors. This work also used the k-means method with morphological segmentation such as erosion and dilation [20].

Quantification of staining intensity in histopathology is relevant only if the absorption of the stain into the preparation is stoichiometric [21]. IHC methods, like most histologic stains, are non-stoichiometric, making it impossible to accurately measure antigen expression based on DAB stain intensity. In addition, the IHC technique uses a series of amplification steps to visualize the results, making it difficult to control the actual signal intensity associated with the amount of antigen. It should be noted that the DAB chromogen does not follow Lambert-Beer's law, which describes a linear relationship between the concentration of a compound and its absorbance (optical density). The stain acts as a light scatterer with a broad, featureless spectrum resulting in a darkly colored DAB, which has a different spectral shape than a lightly colored DAB [21].

Therefore, it was decided to quantitatively analyze the results by determining only the percentage of DAB stain occupancy on the slide relative to the entire surface of the tissue.

As with every method, our solution also has drawbacks. The method does not work well when the plastic culture medium joins the slide and is additionally stained with DAB, or its color does not resemble the synthetic substrate. Moreover, the algorithm may also take into account brown artifacts during segmentation. Another limitation is the MATLAB implementation. A likely gain could be achieved by implementing the proposed algorithm in a more commercial programming language. Lastly, the comparison of the methods was based only on visual evaluation by the expert. More precise methods would be needed to evaluate the performance of the tested algorithms accurately. To obtain automatic matching or mismatching of localized DAB-stained areas, it would have been necessary to compare them with ground truth results (segmentation based on manual labeling), which were not provided in our case.

## Conclusions

The bioinformatics tool developed in the presented work enabled quantitative analysis of marker proteins and assisted the analysis of their spatial distribution by experts. These results might be useful for a better description of structural changes in the reconstructed epidermis and for staining analysis in bioptates, as biopsy material taken by pathologists for diagnostic purposes also has an elongated shape and undergoes IHC analysis. At the same time, it may inspire other researchers to fill the gap in the literature regarding the detection of markers of human epidermal differentiation on IHC images and may be further developed or adjusted to the other research problems in digital pathology.

## Abbreviations

*IHC:* Immunohistochemistry

*HRP:* Horseradish peroxidase

*DAB:* 3,3'-Diaminobenzidine

*NRMB:* Normalized red minus blue

*SE:* Structuring element

*AP:* Average proportion

## Ethics approval and consent to participate

Not applicable

## Consent for publication

Not applicable

## Availability of data and materials

The datasets analyzed during the current study are available in the GitHub repository, https://github.com/DawZam/EpidermaQuant/tree/main/DATA.

## Competing interests

The authors declare that they have no competing interests.

## Funding

This work was founded by the Silesian University of Technology grant number 02/070/BK_23/0043 for Support and Development of Research Potential (MM), by Ministry of Science and Higher Education 'Implementation Doctorate' grant number DWD/7/0396/2023 (DZ), and by National Science Centre, Poland, grant number 2017/25/B/NZ4/01550 (DS).

## Authors' contributions

AG and DS provided the data. DZ designed the algorithm, and the computational framework, and analyzed the data with support from MM. DZ and MM wrote the manuscript. DS conceived the original idea. MM supervised the project. All authors discussed the results and reviewed the manuscript.

## Acknowledgments

Not applicable

# References

[1] S. Magaki, S. A. Hojat, B. Wei, A. So, and W. H. Yong, "An Introduction to the Performance of Immunohistochemistry," *Methods Mol. Biol. Clifton NJ*, vol. 1897, pp. 289–298, 2019, doi: 10.1007/978-1-4939-8935-5_25.

[2] S. W. Jahn, M. Plass, and F. Moinfar, "Digital Pathology: Advantages, Limitations and Emerging Perspectives," *J. Clin. Med.*, vol. 9, no. 11, p. 3697, Nov. 2020, doi: 10.3390/jcm9113697.

[3] M. N. Gurcan, L. E. Boucheron, A. Can, A. Madabhushi, N. M. Rajpoot, and B. Yener, "Histopathological Image Analysis: A Review," *IEEE Rev. Biomed. Eng.*, vol. 2, pp. 147–171, 2009, doi: 10.1109/RBME.2009.2034865.

[4] L. Roszkowiak *et al.*, "System for quantitative evaluation of DAB&H-stained breast cancer biopsy digital images (CHISEL)," *Sci. Rep.*, vol. 11, no. 1, Art. no. 1, Apr. 2021, doi: 10.1038/s41598-021-88611-y.

[5] A. D. Belsare, "Histopathological Image Analysis Using Image Processing Techniques: An Overview," *Signal Image Process. Int. J.*, vol. 3, no. 4, pp. 23–36, Aug. 2012, doi: 10.5121/sipij.2012.3403.

[6] S. Patel, S. Fridovich-Keil, S. A. Rasmussen, and J. L. Fridovich-Keil, "DAB-quant: An open-source digital system for quantifying immunohistochemical staining with 3,3'-diaminobenzidine (DAB)," *PLOS ONE*, vol. 17, no. 7, p. e0271593, Jul. 2022, doi: 10.1371/journal.pone.0271593.

[7] J. Bencze *et al.*, "Comparison of Semi-Quantitative Scoring and Artificial Intelligence Aided Digital Image Analysis of Chromogenic Immunohistochemistry," *Biomolecules*, vol. 12, no. 1, Art. no. 1, Jan. 2022, doi: 10.3390/biom12010019.

[8] R. E. Sziva *et al.*, "Accurate Quantitative Histomorphometric-Mathematical Image Analysis Methodology of Rodent Testicular Tissue and Its Possible Future Research Perspectives in Andrology and Reproductive Medicine," *Life*, vol. 12, no. 2, Art. no. 2, Feb. 2022, doi: 10.3390/life12020189.

[9] S. E. A. Raza *et al.*, "Deconvolving Convolutional Neural Network for Cell Detection," in *2019 IEEE 16th International Symposium on Biomedical Imaging (ISBI 2019)*, Apr. 2019, pp. 891–894. doi: 10.1109/ISBI.2019.8759333.

[10] Z. Swiderska-Chadaj *et al.*, "Learning to detect lymphocytes in immunohistochemistry with deep learning," *Med. Image Anal.*, vol. 58, p. 101547, Dec. 2019, doi: 10.1016/j.media.2019.101547.

[11] P. Bankhead *et al.*, "QuPath: Open source software for digital pathology image analysis," *Sci. Rep.*, vol. 7, no. 1, p. 16878, Dec. 2017, doi: 10.1038/s41598-017-17204-5.

[12] C. A. Schneider, W. S. Rasband, and K. W. Eliceiri, "NIH Image to ImageJ: 25 years of image analysis," *Nat. Methods*, vol. 9, no. 7, pp. 671–675, Jul. 2012, doi: 10.1038/nmeth.2089.

[13] A. Gogler-Pigłowska *et al.*, "Novel role for the testis-enriched HSPA2 protein in regulating epidermal keratinocyte differentiation," *J. Cell. Physiol.*, vol. 233, no. 3, pp. 2629–2644, Mar. 2018, doi: 10.1002/jcp.26142.

[14] S. M. Hewitt, D. G. Baskin, C. W. Frevert, W. L. Stahl, and E. Rosa-Molinar, "Controls for Immunohistochemistry," *J. Histochem. Cytochem.*, vol. 62, no. 10, pp. 693–697, Oct. 2014, doi: 10.1369/0022155414545224.

[15] A. M. Khan, N. Rajpoot, D. Treanor, and D. Magee, "A Nonlinear Mapping Approach to Stain Normalization in Digital Histopathology Images Using Image-Specific Color Deconvolution," *IEEE Trans. Biomed. Eng.*, vol. 61, no. 6, pp. 1729–1738, Jun. 2014, doi: 10.1109/TBME.2014.2303294.

[16] P. Haub and T. Meckel, "A Model based Survey of Colour Deconvolution in Diagnostic Brightfield Microscopy: Error Estimation and Spectral Consideration," *Sci. Rep.*, vol. 5, no. 1, Art. no. 1, Jul. 2015, doi: 10.1038/srep12096.

[17] A. C. Ruifrok, D. A. Johnston, "Quantification of histochemical staining by color deconvolution," *Anal. Quant. Cytol. Histol.*, 23(4):291-9, Aug. 2001, PMID: 11531144.

[18] A. Shehata, S. Mohammad, M. Abdallah, and M. Ragab, "A Survey on Hough Transform, Theory, Techniques and Applications," Feb. 2015.

[19] A. Ashabi, S. B. Sahibuddin, and M. Salkhordeh Haghighi, "The Systematic Review of K-Means Clustering Algorithm," in *2020 The 9th International Conference on Networks, Communication and Computing*, Tokyo Japan: ACM, Dec. 2020, pp. 13–18. doi: 10.1145/3447654.3447657.

[20] V. S and M. R. Shijitha, "Deep Learning based Approach for Efficient Segmentation and Classification using VGGNet 16 for Tissue Analysis to Predict Colorectal Cancer," *Ann. Romanian Soc. Cell Biol.*, pp. 4002–4013, May 2021.

[21] C. M. van der Loos, "Multiple Immunoenzyme Staining: Methods and Visualizations for the Observation With Spectral Imaging," *J. Histochem. Cytochem.*, vol. 56, no. 4, pp. 313–328, Apr. 2008, doi: 10.1369/jhc.2007.950170.